%Paper: hep-th/9302080
%From: andy@nsfitp.itp.ucsb.edu (Andrew Strominger)
%Date: Wed, 17 Feb 93 14:45:21 PST

% jnl.tex (available from hepth) is needed to tex this paper. There is a
%postscript figure appended at the end of the manuscript, which
%should be peeled off and printed. The figure is
%essential for understanding the paper.

\input jnl

\def\a{\alpha}
\def\be{{\beta}}
\def\ga{{\gamma}}
\def\de{{\delta}}

\def\si{{\sigma}}

\def\D{{\Delta}}
\def\Om{{\Omega}}
\def\part{{\partial}}

\rightline{{NSFITP-93-15}}
\rightline{CALT-68-1851}
\rightline{hep-th/9302080}
\title INFORMATION CONSUMPTION BY REISSNER-NORDSTROM BLACK HOLES
%ARE INFORMATION GLUTTONS
\smallskip
\author Andrew Strominger
\affil Institute for Theoretical Physics

and

\affil Department of Physics
University of California
Santa Barbara, CA 93106
 andy@denali.physics.ucsb.edu

and

\author Sandip P.  Trivedi

\affil Department of Physics
California Institute of Technology
Pasadena, CA  91125
trivedi@theory3.caltech.edu

\abstract{
The low-energy scattering of charged fermions by extremal magnetic
Reissner-Nordstrom black holes is analyzed in the large-$N$ and $S$-wave
approximations. It is shown that (in these approximations) information is
carried into a causally inaccessible region of spacetime, and thereby
effectively lost.
It is also shown that there is an infinite degeneracy of quantum black hole
ground states, or ``remnants", which store --- but will not reveal --- the
information. A notable feature of the analysis --- not shared by recent
analyses of dilatonic black holes --- is that the key physical  questions can
be
 answered within the weak coupling domain. We regard these results
as strong evidence that effective information loss occurs in our universe.
}

\endpage
{\bf INTRODUCTION}

Extremal black holes provide a simple laboratory in which to study quantum
mechanical aspects of black holes. There are three general possibilities which
have been discussed for the outcome of a scattering experiment in which a
particle is sent into an extremal black hole and Hawking re-emitted:

\item{I)} The scattering is unitary, with a finite number of quantum states for
the black hole.

\item{II)} The scattering is unitary with an infinite number of asymptotic
quantum states of the black hole, or ``remnants".

\item{III)} The scattering is not unitary, and information is destroyed.

Extensive analyses of extremal black holes in dilaton gravity
at large $N$ over the last year
[\cite{CGHS}] show no evidence that possibility (I) might be
realized, while recent work [\cite{bos}] has shown that possibility (II) and
(III) are much less distinct than previously suspected.

One feature of the large-$N$ analysis of
dilatonic black holes has been in some ways
disappointing: Gravitational collapse inevitably leads to a singularity at
which the large-$N$ approximation breaks down.
Fortunately some key physical questions are not
affected by this breakdown. For example possibility (I) can still be
ruled out at large $N$. However, one can not determine which
of  possibilities (II) or (III) are
realized without solving a strongly coupled quantum problem.

We were thus motivated to search for a model in which possibility (II) is
demonstrably realized at weak coupling\footnote{\dag}{See [\cite{bol}] for
related efforts.}.
After running around in several circles
 we realized that such a model was under our
noses: real-world magnetic Reissner-Nordstrom black holes. Although the
structure of their extremal ground state is much more complex than that of
their
dilatonic cousins, they have two big advantages: large-$N$ can tame their
dynamics, and they exist as solutions to the
Einstein-Maxwell equations, without the introduction of
unobserved fields such as a dilaton.

We wish to study long-wavelength scattering of $S$-wave charged fermions by an
extremal Reissner-Nordstrom black hole with a large magnetic charge and radius,
both given by $Q$ (we take $Q>0$). To render this problem tractable, we make
the $S$-wave approximation in which all
higher angular modes are suppressed. Naively one expects that,
at wavelengths large relative to $Q$, this approximation is good. However,
unlike the dilatonic case studied in [\cite{CGHS}], there
are several subtleties[\cite{ted}]
which have so far prevented a careful justification of the approximation, and
we cannot be sure that it is valid\footnote{\dag}{One subtlety
is that incoming long-wavelength modes may produce regions of high
curvature either
near the origin or the inner
Cauchy horizon. We shall argue later that these regions
are irrelevant to the issue of effective information loss.
Another subtlety has to do with the fact that the centrifugal barrier seen
by the higher partial waves turns off near the horizon. This means firstly
that there are an
infinite number of short-distance but low-energy
 modes near the horizon. We
believe these should not be present in a long-distance effective theory, but we
do not know
now  to define such a theory in a manner consistent with Lorentz
 invariance
and energy conservation. It also means that there are
long-distance low-energy higher angular momentum modes near the horizon which
might be excited as quantum
fluctuations.  As was discussed in [\cite{trv}] the
tidal forces seen by these modes - unlike the s-wave - blow
up at the horizon, thus once excited they might have consequences which are
unaccounted for in the s-wave approximation.
}.
For $N+1$ flavors of fermions the effective two-dimensional $S$-wave theory is
described by
$$
S={1 \over 2 \pi}\int d^2 \si\sqrt{-g}
 \left[e^{-2\phi}R+2e^{-2\phi} (\nabla\phi)^2+2-2Q^2
e^{2\phi}-{1\over2}\sum^N_{i=1}(\nabla f_i)^2\right].\eqno(actn)
$$
The two-dimensional metric appearing in \(actn) is related to the four
dimensional metric by
$ds^2=g_{\a\be}d\sigma^\a d\sigma^\be + e^{-2\phi}d^2\Om$, with $\a , \be $
ranging over   $(r,t)$. The scalar
field $\phi$ measures the (logarithm of) the area of two spheres of constant
radius. The first four terms in \(actn) arise directly from dimensional
reduction of the four-dimensional Einstein-Maxwell action. The last term arises
from the bosonization of fermion $S$-wave modes studied by Callan and Rubakov
[\cite{crv}] in GUT theories. One charged linear combination of the original
$N+1$ flavors acquires a mass from electromagnetic effects. The dynamics of
this
mode was studied for dilaton black holes in [\cite{alf}], but because of its
mass it decouples at sufficiently low energies. The two-dimensional relic of
the four-dimensional gauge field is suppressed in \(actn), as
it cannot be excited
by the neutral fields and so may be consistently neglected.
Previous work on the model defined in \(actn) and related
models obtained by dimensional reduction can be found in [\cite{prev}]
and [\cite{trv}].

One might hope that for large $Q$
particle-hole scattering could be adequately  analyzed in a
semiclassical loop expansion of the reduced theory, but in fact large $N$ will
be needed in addition to the $S$-wave approximation, for several reasons.
First, as pointed out in [\cite{PSSTW}], the temperature fluctuations of a
near-extremal charged black hole go as
$$
{\D T\over T}\sim \bigl[{\hbar \over \sqrt{M-Q}}\bigr]^{1 \over 2} , \eqno(tfl)
$$
so the leading
 semiclassical formula for the  temperature (and radiation rate)
becomes unreliable very near extremality.
 But in the large-$N$ limit, where $\hbar \to 0$ (while  $N \to \infty $
keeping $ N \hbar $ fixed) we see that
$$
{\D T\over T} \to 0,
$$
so that near-extremal black holes are indeed characterized by a
definite temperature.
%one holds $GN,
%m=M/N$ and $q=Q/N$ fixed. \(tfl) then becomes
%$$
%{\D T\over T} \sim {1\over N(m-q)}\to 0
%$$
%so that near-extremal black holes are characterized by a definite temperature.

A second problem with the loop expansion was discussed in [\cite{trv}].  For a
non-extremal Reissner-Nordstrom black hole, the one-loop contribution to the
expectation value of the stress tensor diverges on the inner (but not the
outer) horizon. This is related to the classical instability of the inner
horizon, as studied in many papers [\cite{hrz}]. This divergence persists,
albeit in a softened form, in the extremal limit in which the two horizons
coalesce. Since the one-loop corrections are divergent, the loop expansion is
clearly unreliable.

Although frightening at first, these divergences are in fact rather benign and
can be controlled within the $1/N$ expansion. The leading large-$N$ equations,
in which one-loop quantum back reaction is included, have solutions which are
in a sense ``near" to corresponding classical solutions. In particular
[\cite{trv}], there is an extremal, zero-temperature ground-state solution
with causal structure identical to that of the classical solution. The
large-$N$ geometry is near to its classical counterpart, but third and higher
derivatives of the fields are divergent near the horizon. We shall see in this
paper that divergences encountered in particle-hole scattering are also benign,
although the behavior of the stress tensor near the horizon leads to an
unexpected non-analytic large-$N$ mass-area relationship, which differs from
the classical result even at large $Q$ sufficiently near extremality.

{\bf CALCULATION}

Previous analyses of large-$N$ two dimensional gravity have been largely
carried out in conformal gauge. This gauge is somewhat awkward in the present
context. For example even the classical solutions are known only implicitly in
this gauge. A more convenient choice is light-cone gauge\footnote{\dag}
{Previous
 light-cone gauge analyses of dilatonic black holes can be
found in [\cite{leg}].},
 for which the two dimensional metric takes the form
$$
ds^2=-h(dv)^2+2drdv,\eqno(lcm)
$$
 $\sqrt{-g}=1$ and the scalar curvature is $R=-\part^2_rh$.
 In this gauge one
can solve the classical equations and obtain an analytic expression --- known
as the Vaidya metric --- for
arbitrary null infalling matter, characterized by a stress tensor obeying
$T_{ra}=0$:
$$
h=1-{2M(v)\over r}+{Q^2\over r^2}~, \eqno(rdya)
$$
where $2\part_vM=T_{vv}$. The large-$N$ trace and dilaton equations
may then be
written in the form
$$
\square\phi=\part_r\Sigma=\Sigma \part_rU+
{Q^2e^{4\phi}(1+\ga e^{2\phi})\over(1-\ga e^{2\phi})}-\ \
{e^{2\phi}\over1-\ga e^{2\phi}},\eqno(pheq)
$$
$$
R=-\part_r^2h={2\over1-\ga
e^{2\phi}} \left[e^{2\phi}-2Q^2e^{4\phi}-\Sigma\part_r\phi
\right], \eqno(heq)
$$
where
$$\eqalign{
\Sigma & \equiv 2\part_v \phi+h\part_r\phi, \cr
U & \equiv 2\phi-{1\over 2}\ln(1-\ga e^{2\phi})\cr}. \eqno(ueq)
$$
and
$$
\ga= {N\hbar \over 24} \eqno(gamma)
$$
A future (past) apparent horizon is a  zero of $\Sigma\ \ (\part_r\phi)$,
 which implies
$(\nabla\phi)^2=0$.
One important linear combination of the constraint equations is local
$$
\eqalign{
e^{2\phi}l^a l^b T_{ab}^Q & \equiv {1 \over 2} \ga e^{2 \phi} ({1 \over 2} h
\part_r^2h
             -{1 \over 4}(\part_r h)^2+\part_v\part_rh) \cr
& = \part_v\Sigma +{h \over 2} \part_r\Sigma -{1 \over 2}\Sigma^2 -
                  {1 \over 2} \part_rh \Sigma .  \cr  }\eqno(vv)
$$
where the components of the null vector $l$ are $(l^r,l^v)=(h/2,1)$.
Fortunately, the other linear combination, which is non-local,
shall not be needed.
%$(v,v)$ constraint equation is
%$$
%\eqalign{
%e^{-2\phi} & (\part_v\phi\part_v\phi-
%\part_v^2\phi-{1\over 2}\part_v h\part_r\phi
%+{1\over 2} h\part_rh\part_r\phi-{1\over2}\part_rh\part_v\phi)
% +h(e^{2\phi}-4Q^2
%e^{4\phi})\cr
%&=\ga((\part_rh)^2-2h\part_r^2h+4
%\part_r\part_vh)\equiv T^Q_{vv}.\cr}\eqno(vv)
%$$
%The $(r,r)$ constraint is non-local and  fortunately shall not be needed.

The extremal solutions were studied in [\cite{trv}] and found to be of  two
kinds referred to therein as the even and odd extensions. Here we
shall focus on
the  odd extension, which reduces to the classical solution as
$\hbar \rightarrow 0$,  and denote it as $\phi_0(r)$ and $h_0(r)$
\footnote{\dag}{The even solutions are in many ways more interesting
since they correspond to spacetimes
free from any malevolent singularities. But their stability and
response to perturbations cannot be studied in the approximations
used here.}.
%The extremal solutions denoted $\phi_0(r)$ and $h_0(r)$, was
%studied in [\cite{trv}].
This solution has a timelike singularity at the
``origin" where $e^{-2\phi_0}=\ga$.
Near the horizon $r_H$, it was shown [\cite{trv}] that
the fields have the non-analytic behavior
$$
\eqalign{
\phi_0-\phi_H &=\be x|x|^\de, \cr
h_0 &=\a_1x^2+\a_2x^3|x|^\de.\cr}\eqno(texp)
$$
Where $\phi_H\equiv \phi(r_H)$,  $x\equiv r-r_H$ and
$$
\delta ={3 \over 2}\bigl[\sqrt{1+{8 \ga \over 3(e^{-2\phi_H} -\ga)}}-1
\bigr],
 \eqno(deltriv)
$$
with
$$
e^{-2 \phi_H} =Q^2\bigl[{1+\sqrt{1+{4 \ga \over Q^2}} \over 2} \bigr].
 \eqno(phih)
$$
$\de$ tends to zero for large $Q$.
To leading order in ${\ga \over Q^2}$
$$
\a_1={1 \over Q^2 }, \eqno(aone)
$$
$$
\a_2 = -{2 \over Q^3}, \eqno(atwo)
$$
$$
\be =-{1 \over Q}, \eqno(dtwo)
$$
$$
r_H=Q,\eqno(rhq)
$$
and
$$
e^{-\phi_H}=Q. \eqno(phq)
$$
While for large $Q$ one can safely use these approximations to
$\a_1,~ \a_2,~ \be,~ r_H $ and $\phi_H$, we do not omit terms subleading
in $1/Q$ in the expression for $\delta$ because such an
approximation would break down very near the horizon (at $x$ less than of
order $Qexp(-Q^2/ \ga)$)\footnote{\dag}{Note that there is an issue of orders
of limits here: we take
$N \rightarrow \infty$ before $\mu \rightarrow 0$.}.
The non-analyticity in \(texp) leads to divergences
for example in the second derivative of the
curvature.

Let us now consider an  incoming matter shock wave\footnote{\dag}{Strictly
 speaking shock waves are not allowed in the
long-distance effective field theory, but the case of a smooth pulse is
qualitatively similar.}
along $v_0$ whose classical stress tensor obeys
$$
l^al^bT^f_{ab}=2\mu\de(v-v_0). \eqno(mvv)
$$
We wish to compute, following [\cite{BDDO}], $\phi$ and $h$
perturbatively in $\mu$ in a
Taylor expansion above the shock wave. $\phi$ is continuous across the shock
wave, while $h$ has a discontinuity which is determined by the constraints and
is classically equal to ${-2 \mu\over r}.\ \ \Sigma$ (defined in \(ueq))
also has a
discontinuity $\de\Sigma$ across $v_0$, which, according to \(pheq), obeys
$$
\part_r (e^{-U} \de\Sigma)=0. \eqno(dseg)
$$
The integration constant is determined from the asymptotic boundary condition
$\de\Sigma\to{2 \mu\over r^2}$. One thereby obtains
$$
\de\Sigma=2\mu e^U.\eqno(mgu)
$$
Near the horizon $r_H,\Sigma_0$ (the value of $\Sigma$
 below the shock wave) has a
higher order zero:
$$
\Sigma_0\approx\a_1 \be(1+\de)x^2|x|^\de .   \eqno(hoz)
$$
Since $\de\Sigma$
 is non-zero at $r_H$, this zero is split into two simple zeros
which are (by definition) the inner and outer apparent horizons, as illustrated
in the figure. To leading order in ${\ga\over Q^2}$ and ${\mu
\over Q}$  one finds that the
locations of the horizons are given by
$$
r_\pm\approx r_H\pm Q\  ({2 \mu \over Q })^{{1\over2+\de}}.\eqno(rpm)
$$
%To this order, $\mu$ is the deviation of the black hole mass from its extremal
%value.
Comparing with \(texp) and recalling that the dilaton measures the area
of the two spheres, we see that the area $A_H$ of the outer horizon obeys
$$
A_H-A_0\simeq 8\pi \ Q^2 \ ({2 \mu \over Q})^{{1+\de\over2+\de}},\eqno(mar)
$$
where $A_0$ is the extremal area. Thus the mass-area relation
is non-analytic. Notice that no matter how small $\de$
is, there is always some value of $\mu~~(\mu\sim Qe^{-{1\over\de}})$ below
which
\(mar) is not well approximated by the classical relation
$A_H-A_0\sim 8\pi \sqrt{ 2 \mu Q^3}$.

The mass of the black hole will of course decrease due to Hawking radiation
and we expect it to settle back to extremality.
To study this, we first calculate the trajectories of the two  apparent
horizons, denoted by $ \hat r_\pm$. We again work to leading order in
${\ga \over Q^2}$ and ${\mu \over Q}$.
Since $\Sigma$ vanishes along $\hat r_\pm$ one has
$$
\part_v \Sigma=-\part_r\Sigma \part_v\hat r_\pm. \eqno(slope)
$$
Now \(mgu) and \(hoz) imply that
$$
\part_r \Sigma(r_\pm) \simeq \mp
                          {2 \over Q^2} \ ({2 \mu \over Q})^{1+ \delta \over
                                           2 + \delta}. \eqno(dersx)
$$
Similarly \(vv) implies that at $r_\pm$
$$
\eqalign{
{h \over 2} \part_r \Sigma + \part_v \Sigma =
 {\ga e^{2 \phi} \over 2} ({h_0 \over 2} \part_r^2 \de h
                              +{\de h \over 2} \part_r^2 h_0
                              -{1 \over2} \part_r h_0 \part_r \de
h+\part_v\part_r \de h) +
e^{2 \phi}l^al^bT_{0ab}^Q.\cr}
                               \eqno(vvtwo)
$$
It turns out that the dominant contribution for small $\mu$ is given by the
second term on the right hand
side, which involves $\de h$ with no derivatives.
To evaluate this term we  need to know $\de h$ just above the shock wave
which, according to
\(heq) obeys
%Just above the shock wave the decrease is given by
%$$
%\part_v\mu(v_0)=T^Q_{vv}(r_+,v_0)\eqno(dmt)
%$$
%given in \(vv).
%To leading order in $\mu$
%$$
%T^Q_{vv}={1\over4}\part_rh_0\part_r\de h-{1\over4}h_0\part_r^2\de
%h-{1\over4} \de h\part_r^2 h_0+{1\over2}\part_r\part_v\de
%%h+T^Q_{0vv}\eqno(thh)
%$$
%evaluated at $(r_+,v_0)$.
% The last two terms are non-analytic in $\mu$ but can be
%seen to be subleading. To evaluate the first three terms we need to know $\de
%h$ just above the shock wave which, according to \(heq) obeys
$$
\part_r^2\de h={2\part_r\phi\de\Sigma\over1-\ga e^{2\phi}}, \eqno(rrh)
$$
so that
$$
\de h=4\mu\int dr {1\over\ga} (-1+{1 \over \sqrt{1-\ga e^{2\phi}}}).
\eqno(dhm)
$$
(The integration constants are fixed by the requirement that $\de h$
asymptotically
vanish.) Furthermore, to leading order  $h \part_r \Sigma $ vanishes so
that \(vvtwo) reduces to
$$
\part_v \Sigma(r_{\pm}) \simeq  - {\ga \mu \over Q^5} .
                                            \eqno(sigv)
$$
\(slope) and \(rpm) then imply that right above the shock wave
$$
\part_v \hat r_\pm \simeq - {\ga (\hat r_\pm -r_H)\over 4 Q^2}\
                                . \eqno(sltwo)
$$
%One may now evaluate \(dmt) as
%$$
%\part_v\mu=C\mu \eqno(trm)
%$$
%where
%$$
%C={\ga\over Q^2}(1+O({\ga\over Q^2})).
%$$
%In the adiabatic approximation
% the mass then decays
%exponentially back to its extremal value
%$$
%\mu(v)=\mu_0 e^{-\ga (v-v_0)/Q^2}. \eqno(mxp)
%$$
%We expect the adiabatic approximation to be good for large
%$Q^2$, but have been unable to carefully justify this.
%The trajectories of  the two apparent horizons, denoted $\hat r_\pm(v)$, are
%simply
%determined from $T^Q_{vv}$. Since $\Sigma$ vanishes along $\hat r_\pm$ one has
%$$
%\part_v \Sigma=-\part_r\Sigma \part_v\hat r_\pm
%$$
%It then follows from the equations of motion and constraints that
%$$
%\part_v\hat r_\pm = e^{2\phi} {(1-\ga e^{2\phi})(2T^Q_{vv}+hT^Q_{rv})\over
%2Q^2 e^{4\phi}(1+\ga e^{2\phi})-e^{2\phi}} \approx \ga{\hat r_\pm\over Q^2}.
%$$
%where $T^Q_{rv}=\ga\part_r^2 h$.

   To proceed further,
we evoke the adiabatic approximation in which the black hole
is taken to  evolve slowly
so that it's dynamic geometry may be approximated by
a sequence of static ones. We expect the adiabatic
approximation to be good for large
$Q^2$, but have been unable to carefully justify this. In this
approximation \(sltwo) continues to hold
everywhere along $\hat r_{\pm}$.
Thus the inner horizon moves out towards $r_H$ while the outer horizon moves
in towards  $r_H$, and the black hole exponentially approaches its extremal
ground state. Note however that while the black hole
is excited, the trajectories
of $\phi=\phi_H$ and $r=r_H$ are spacelike.
The event horizon therefore is shifted outward (relative to the original
apparent horizon) by the
scattering process, as illustrated in the figure.

 Also, within the adiabatic approximation \(rpm)  relates the position of
the outer horizon $r_+$ to $\mu$.  \(sltwo) then implies that

$$
\part_v \mu \simeq  -{\ga \over 2 Q^3} \mu,  \eqno(rmls)
$$
so that  the mass decays exponentially back to it's extremal value as
$$
\mu(v) \simeq \mu e^{{-\ga (v-v_0) \over 2 Q^3}}. \eqno(mxp)
$$

There are two regions in which the large $N$ equations used here cannot be
trusted. The first is near the origin $e^{2\phi}=\ga$, where the curvature
becomes large and higher-dimension corrections to the Einstein-Maxwell theory
are important. The second is the future Cauchy horizon, or the extension of
${\cal I}^+$ inside the event horizon.
 An observer inside the black hole crosses this
Cauchy horizon in finite time, yet is able to see all of the universe outside
the black hole before doing so. There is therefore a large energy concentration
near this surface, the effects of which are subtle and have been analyzed in
many papers [\cite{hrz}].

Fortunately, physics {\it outside\/} the horizon is insensitive to the
(intractable) behavior of the system in these regions. To see this, consider a
Hamiltonian $H$ which evolves along the series of spacelike slices asymptotic
to the
slice $\Sigma = \Sigma_I \cup {\cal I}^+$ where, as depicted in
the figure, $\Sigma_I$ is a spacelike surface inside the horizon.
These slices can be chosen to completely cover the spacetime outside the
horizon.
$\Sigma_I$ can be chosen so that it
avoids the difficult region near the future Cauchy horizon, and so that the
intersection of the shock wave
with $\Sigma_I$ is in weak coupling. Although $\Sigma_I$ extends into the
strong coupling region near
$e^{2\phi}=\ga$, this does not present any difficulties because the
system is unexcited in that region. The non-trivial dynamics are everywhere
weakly coupled
for all time, and our approximations should be valid.

{\bf DISCUSSION}

We now argue that our results imply that an
arbitrarily large amount of information can be sent into the black hole,
and will
never emerge again in the universe from which it was  thrown in.
The black hole relaxes to extremality with a characteristic time
$$
t_c={Q^3 \over \ga}. \eqno(tchr)
$$
Consider experiments in which an arbitrarily large
number of
wavepackets are sent in from  ${\cal I}^-$ spaced at intervals of
$t_c$ seconds. In the process, an arbitrarily large amount of
information is sent in. In order for no information loss to occur,
in the asymptotic future, all correlations
between the state inside and the state outside the horizon should be
destroyed. This can occur only if the state inside is unique and independent
of the initial state of the infalling matter.
The preceding analysis shows that nothing catastrophic happens to the
infalling matter as it crosses the apparent horizon
so
in the asymptotic future the
state inside the event horizon (on $\Sigma_I$) will depend heavily on the
incoming scattering state.
Indeed, since the system is still weakly coupled on $\Sigma_I$,
the quantum state of the left-moving
conformal $f$-matter will be essentially the same as on ${\cal I}^-$.
Thus in the course of this experiment an
arbitrarily large amount of information will be carried into
the causally inaccessible region inside the event horizon and thereby
be effectively lost.

It is also evident that, with respect to the  time slicing described above,
this
is a theory with an infinite number of remnants. What we mean by
this statement is that there are an infinite number of solutions of the
large-$N$ constraint equations on a spacelike slice
which are identical outside the horizon,
but have differing $f$-matter configurations inside the horizon,
corresponding to an infinite degeneracy of
large-$N$ semiclassical quantum
states\footnote{\dag}{Actually, if we enforce the constraint
that the incoming matter excitations are long wavelength on ${\cal I}^-$
- in accord with our approximations-this would not be the case because
the incoming matter excitations must be late enough so that they are still
in weak coupling when they arrive at $\Sigma$, yet early enough to avoid
a potential pile up of energy density near the
future Cauchy horizon. There are
only a finite number of states in this finite interval above  any
given wavelength. This problem can be avoided by choosing a different surface
$\Sigma_H$ defined as the (spacelike or null) surface along which
$\phi$ takes the constant value $\phi_H$ characterizing the
horizon of an unperturbed extremal solution. This is a geodesically
complete surface which is everywhere in weak coupling. Since any
finite point on $\Sigma_H$ is an infinite distance from $i^+$,
there is clearly
no pile up of energy density. Furthermore, because the event horizon is moved
out by each
scattering process, this surface is well behind the event horizon if
many $f$-particles (and much information) are thrown in to the black hole.
The potentially infinite amount of information on
$\Sigma_H$ is therefore unavailable to
an observer on ${\cal I}^+$.}.

However it is important to note that the interpretation that the
information is stored in these remnants may be dependent on the
slicing. For example we might have chosen the asymptotic interior
surface $\Sigma_I$ so that the $f$-wave arrives at the singularity before
intersecting $\Sigma_I$. In this case it is a logical possibility
that the information is destroyed when it arrives at the singularity,
in which case one would not conclude that the information is stored
inside the black hole\footnote{\dag}{Another possibility is that
boundary conditions might be specified at the timelike singularity
to reflect the matter - and the information - up to
the future Cauchy horizon and possibly on to the next universe.
In this case all the information will be present on
$\Sigma_I$ no matter how it is chosen.}.

Of course, physics outside the event horizon cannot, by causality,
depend on the choice of slicing inside the event horizon, which
we are therefore free to choose for our own convenience.
Consequently the observer outside the horizon cannot possibly distinguish
between actual information loss and storage by remnants. The choice
of slicing made
in this paper was motivated by  the desire to avoid the difficult
dynamics near the singularity, and it is consistent
with this choice to describe
the theory as having an infinite number of remnants.

We should note that although the calculations above were carried out in the
$N \to \infty$ limit all the important conclusions continue to hold when
$N$ (or $Q$) is sufficiently large but finite. The distinction is important to
make
because when $N \to \infty$, $\hbar \to 0$, so that
the Bekenstein-Hawking entropy of the black hole - which goes as ${1 \over
\hbar}$ -
goes to $\infty$.
Thus it might be claimed that our conclusions are simply a consequence of
working in a limit where the ground state is infinitely degenerate and
that at finite
$N$ there would be an upper
limit on the information
the black hole can carry and a finite number of remnants. Finite
$N$ differs from $N \to \infty$ in that we have to keep track of the
quantum fluctuations in
the metric and dilaton, and these might be potentially large close to the
horizon. But the larger $N$ is, the closer one must approach the horizon in
order
for these effects to be significant. Similarly, the adiabatic approximation
would break down for finite $N$ sufficiently close to extremality. But again
for large enough $N$ this occurs only very near extremality.
Thus by sending in energy at a judicious rate, for $N$ large
but finite, one could keep the black hole close enough to extremality
($ {\mu \over Q} $ small enough) for our approximations to hold, but far enough
from extremality for the finite $N$ effects to be insignificant at the apparent
horizons. The black hole would then respond according to the  calculations
above except for the first moments after it departs from extremality and the
last moments before finally settling down. We would thus conclude that
even for finite $N$ (when the entropy is finite) that the black hole can
consume an arbitrarily large amount of information and store it in an
infinite number of remnant states.

Finally, one may be concerned that
this  infinite degeneracy of states will lead to divergent black
hole pair production rates. In fact magnetic black hole pair production was
computed semiclassically in [\cite{gast}]
 and found to be finite. The reason for this was
discussed at length in [\cite{bos}] : extremal black holes do not behave
quantum mechanically like elementary particles.

Thus we have found a system which can be seen --- without resorting to
speculations about strong coupling dynamics --- to solve the information puzzle
by storing it in an infinite degeneracy of black hole quantum states.
Further this two-dimensional system might be a good approximation to real-world
long-wavelength
fermion-magnetic black hole scattering.

{\bf Acknowledgements}

We wish to thank T.~Banks, J.~Bekenstein, S.~Giddings, J.~Hartle, G.~Horowitz,
M.~O`Loughlin, J.~Preskill and L.~Thorlacius for useful discussions. This work
 is supported in part by DOE grant No. DEAC-03-8ER4050,
and by the National Science Foundation
under Grant No. PHY89-04035.

\head{References}

\refis{hrz} J.M. McNamara, \journal Proc. R. Soc. London, A358, 449, 1978; {\bf
A364}, 121 (1978); Y. G\"ursel, V.D. Sandberg, I.D. Novikov, and A.A.
Starobinsky, \pr D19, 413, 1979; R.A. Matzner, N. Zamorano, and V.D. Sandberg,
{\it ibid.\/} {\bf 19}, 2821 (1979); S. Chandresekhar and J.B. Hartle, \journal
Proc. R. Soc. London, A284, 301, 1982; N. Zamorano, \pr D26, 2564, 1982; E.
Poisson and W. Israel, \prl 63, 1663, 1989; \pl B233, 74, 1989; \pr D41, 1796,
1990; A. Ori, \prl 67, 789, 1992.

\refis{gast} D. Garfinkle and A. Strominger ``Semiclassical Wheeler Wormhole
Production" \pl B256, 146, 1991.

\refis{leg} H. Terao ``Two-Dimensional Black Hole Evaporation in Light-Cone
Gauge" Kanazawa preprint 92-19 (1992). X. Shen ``Quantum Dilaton graity in
Light-Cone Gauge" Cern preprint TH-6633 (1992).

\refis{CGHS} C.G. Callan, S.B. Giddings, J.A. Harvey and A. Strominger,
``Evanescent Black Holes", \pr D45, R1005, 1992; For recent reviews see J.A.
Harvey and A. Strominger,
``Quantum Aspects of Black holes" preprint EFI-92-41, hep-th@xxx/9209055, to
appear in the proceedings of the 1992 TASI Summer School in Boulder, Colorado,
and S.B. Giddings, ``Toy Models for Black Hole Evaporation" preprint
UCSBTH-92-36, hep-th@xxx/9209113, to appear in the proceedings of the
International Workshop of Theoreticl Physics, 6th Session, June 1992, Erice,
Italy.

\refis{BDDO}T. Banks, A. Dabholkar, M.R. Douglas and M. O'Loughlin, ``Are
Hormed Particles the Climax of Hawking Evaporation?", \pr D45, 367, 1992; J.G.
Russo, L. Susskind and L. Tholacius, ``Black Hole Evaporation in $1+1$
Dimensions", \pl B292, 13, 1992.

\refis{PSSTW} J. Preskill, P. Schwarz, A. Shapere, S. Trivedi and F. Wilczek,
\journal Mod. Phys. Lett., A6, 2353, 1991.

\refis{bol} T. Banks and M. O'Loughlin ``Nonsingular Lagrangians for
Two Dimensional Black Holes" Rutgers preprint RU-92-61 (1992).

\refis{bos} T. Banks, M. O'Loughlin and A. Strominger ``Black Hole Remnants and
the Information Puzzle" hep-th/9211030, {\sl Phys.\ Rev.\ D\ } to appear.

\refis{prev} M. McGuigan, C. R. Nappi and S. A. Yost, ``Charged Black Holes
In Two Dimensional String Theory'' IASSNS-HEP-91/57; O. Lechtenfeld and
C. R. Nappi, ``Dilaton Gravity and No Hair Theorem in Two Dimensions ''
IASSNA-HEP-92-22; D.A. Lowe, ``Semiclassical Approach to Black Hole
Evaporation '' PUPT-1340.

\refis{trv} S. Trivedi ``Semiclassical Extremal Black Holes'' Caltech preprint
CALT-68-1833 (1992).

\refis{alf} M. Alford and A. Strominger ''S-Wave Scattering of Charged
Fermions by a Magnetic Black Hole''  {\sl Phys.\ Rev.\ Lett\ } 69, 563, 1992.

\refis {crv} C.~Callan, Phys. Rev. D25 (1982) 2141; Phys. Rev.
D26 (1982) 2058; Nucl. Phys. B212 (1983) 391;
V. Rubakov, Pis'ma Zh. Eksp. Teor. Fiz. 33 (1981) 658 (JETP
 Lett. 33 (1981) 644); Nucl. Phys. B203 (1982) 311.

\refis{ted}T. Jacobson ``Black Hole Evaporation and Ultrashort Distances''
 {\sl Phys.\ Rev.\ D\ }44 (1991) 173; ``Black Hole Radiation in the
Presence of a Short-Distance Cutoff'' UMDGR93-32, ITP preprint (1993).

\endreferences

\head{Figure Caption}

\noindent Figure 1. A shock wave incident on an extremal Reissner-Nordstrom
black hole splits the apparent horizon $r_H$ into a pair of apparent horizons
$r_\pm$, which then exponentially decays back to $r_H$. The even horizon is
outside $r_H$. The asymptotic spacelike surface $\Sigma$ is positioned so that
the shock wave intersects it at large radius and  weak coupling, and it avoids
the Cauchy horizon.

\end